# A Brief Survey on Interactive Automotive UI

Gowdham Prabhakar & Pradipta Biswas*
Indian Institute of Science
*pradipta@iisc.ac.in

**Abstract:** *Automotive User Interface (AutoUI) is relatively a new discipline in the context of both Transportation Engineering and Human Machine Interaction (HMI). It covers various HMI aspects both inside and outside vehicle ranging from operating the vehicle itself, undertaking various secondary tasks, driver behaviour analysis, cognitive load estimation and so on. This review paper discusses various interactive HMI inside a vehicle used for undertaking secondary tasks. We divided recent HMIs through four sections on virtual touch interfaces, wearable devices, speech recognition and non-visual interfaces and eye gaze controlled systems. Finally, we summarized advantages and disadvantages of various technologies.*

## 1. Introduction

In 1886, two German engineers, Benz and Daimler, started the first automobile company known as Mercedes-Benz. Daimler worked on internal combustion (IC) engine, and initially, he tested it on a bicycle while Benz designed a vehicle that was patented in 1886. Daimler's vehicle appeared a few months later. During a similar time in North America, George B. Selden patented his automobile engine and its application in a car in 1895 while Henry Ford introduced the iconic Model T in 1911. In 1926, two-third of running cars were Model T. It may be noted that although the first automotive was patented in late 18th century, it is not until 1924, Kelly's Motors installed the car radio, which can be considered as the first secondary task allowed to drivers in addition to driving. In 1939, Packard introduced the first air-conditioned car. In 1960, the tape music players were installed in cars. The Compact Disk (CD) player was introduced in cars during 1980. MP3 players started ruling the in-car music systems from 2000 till date. Pioneer introduced the smartphone-based head unit with iOS support in 2011. In 2012, Tesla introduced the full size single digital display for secondary tasks. Early cars like Model T had minimal controls to operate the vehicle. As the in-car features were increasing with the introduction of music player and air conditioner like Packard, the user interface (UI) to control the functionalities became complex. The digital interfaces in Tesla cars offered the users to personalize interaction in terms of the hierarchy of icons displayed. However, the single large display may introduce long time to search for appropriate screen element and in turn, increase the duration of eyes-off-road. This increase in complexity of car dashboards requires researchers in improving the UI Early stages of automotive User Interface (UI) used analog interfaces consisting of buttons and switches, which require physical touch to manipulate. In physical touch based interfaces [Schnelle 2019], either the physical control needed to be visually searched or the feedback of selection needed to be visually confirmed. Although the visual search or feedback could be avoided with practice but it often distracted drivers from their visual attention in driving. As physical buttons cannot adapt to dynamically varying content, digital screens with GUI (Graphical User Interface) took over the market. These GUI-based systems accommodated new functionalities in addition to existing features and later gained popularity in the name of IVIS (In-Vehicle Infotainment System). Introduction of new functionalities in IVIS affects attention and learnability of drivers to interact. NHTSA reported 17% of car crashes due to operating such systems while driving [NHTSA 2012]. Though researchers are studying different input modalities, reducing cognitive load, visual, and physical effort to operate IVIS needs further investigation.

Kern (2009) classified five main interaction areas in a car – Windshield, Dashboard, Centre Stack, Steering Wheel, Floor and Periphery. Existing Graphical User Interface (GUI) based IVIS either has a Head-Down Display (HDD) at Centre Stack or Dashboard where the driver looks down to operate or a Head-Up Display (HUD) at Windshield, where the driver looks up to operate. They are operated using modalities like physical buttons, touchscreen, and voice recognition systems. Researchers have investigated hand gesture tracking [Ohn-Bar 2014, May 2014, Ahmad 2016, Prabhakar 2016], haptic feedback [Chang 2011, Vito 2019], personalizing displays to help drivers in parking vehicles [Feld 2013, Normark 2015].

This paper presents a brief review of interactive Human Machine Interfaces (HMI). It may be noted that a plethora of research in Automotive UI concerns about distraction detection, cognitive load estimation and drivers' behaviour analysis. This paper does not review such technologies – rather only covered interactive technologies used inside a car to operate various controls related to driving, communication and entertainment. The next section looks back historical HMI and their evolution through different functions. Subsequent sections described the state-of-the-art through the following four sub-headings

- **Virtual touch interface** covers research on interaction techniques used to operate existing touchscreen displays without physical touch
- **Wearable devices** discuses interfaces requiring drivers to physically wear a sensor
- **Speech recognition and non-visual interfaces** cover interfaces that do not require drivers to look at the display, rather exploits non-visual modalities like voice, tactile and haptic systems
- **Gaze controlled interfaces** describe systems that utilizes drivers' glance to directly manipulate displays and do not require them to take hands off control.

## 2. Early automotive HMI

In the early stages, HMI in automotive context focused on providing drivers with information related to primary driving parameters like speed, throttle, and revolution counter. As driving duration increased, features like music players got integrated into cars leading to an increase in complexity of HMIs. The combined provision of information with entertainment functionalities led to the revolution of infotainment systems [Bosshart, 2009]. In addition to vehicle-related information, current infotainment systems are integrated with advanced driver assistance and entertainment features. The dashboard of Ford S-Max features entertainment, navigation and phone options in the middle console, and trip computer information and safety functionalities in the display behind the steering wheel. The usability of such HMI systems attracted attention of researchers due to their increased complexity [Ariza, 2007]. The latest HMIs incorporate a GUI which can be operated through touch, speech, and gesture recognition systems. Interfaces like motorised mirrors which are electrically operated provide more flexibility than mechanical mirrors in terms of adjusting the field of view and eliminating blind spots. As new functionalities also add more complexity to human-machine interaction, designing an HMI to accommodate latest functionalities require an understanding of different end-users' requirements, including first-time users. A brief history of integrating different functionalities like music player, navigation and telephone into car HMIs are discussed in the following sub-sections.

**Music:** The first in-car radio was launched in Ford Model T during 1954, and car radios were only available as accessories to a few cars. It was not a standard even in high-quality cars like Mercedes-Benz 300 SL. In 1956, a record player named "Highway Hi-Fi" was offered in vehicles like Chrysler, DeSoto, Dodge and Plymouth. In 1968, Philips launched their first in-car cassette player enabling drivers to listen to their custom tapes. After the invention of CD, Philips developed their first in-car CD player in 1983. The in-car CD players became popular only during late 1990s due to compatibility with MP3 files. Usage of in-car CD players improved usability of music players as they accommodated feature to change track easier than cassette players and led to a decrease in driver distraction.

**Navigation:** As the number of driveways and number of vehicles on the road increased, real-time information about navigation route and traffic congestion turned important inside car cockpit. Traffic information was first broadcasted via radio during 1970s. In 1990s, Global Positioning System (GPS) navigation was facilitated inside cars by displaying a visualisation on Liquid Crystal Display (LCD) screens [Bellis]. The screen inside the car cockpit required wide area and rotary push buttons for interaction to accommodate visual information about navigation. The navigation system further required data processing, GPS module and different sensors for precisely locating the vehicle.

**Telephone:** The first telephone was installed in a car by Lars Magnus Ericsson in 1910, where the telephone could be electrically connected to telephone poles installed along the road [Wheen, 2010]. A car radiophone service network was used in 1970s. In 1982, mobile telephone service by Nordic Mobile Telephone was the mainstream service for automotive. As cell phones became affordable and popular, car phones became obsolete. In 2001, the Bluetooth special interest group introduced a hands-free car kit for mobile phones with speech recognition system.

## 3. Virtual touch interface

Virtual touch systems aim to speed up the interaction with touchscreen system as they can activate a touchscreen without physically touching it. This paper investigated three types of technologies for virtual touch systems - infrared (IR), wearable device, and inertial measurement unit (IMU) based systems. The infrared sensors track the position of hand or fingers and have limitations in terms of field of view, accuracy, and latency of tracking in different vibrating and lighting conditions inside a car [Biswas 2017b]. The wearable systems using sensors attached to hand and fingers are also investigated for performing secondary tasks like wearable physiological sensors for music recommendation [Ayata 2018]. The third kind of virtual touch system is based on existing remote control-based pointers [Witkowski 2014], as illustrated by Witkowski. Khan has worked on a handheld device to interact with smart homes [Khan 2018]. Such systems are prone to unintended activation of functions in the infotainment system due to uncontrolled tracking of actions from fingers, hands, and remote even when the driver does not intend to operate the system.

A set of virtual touch interfaces used AI based target prediction technologies. Pointing target prediction technologies were already investigated for human-computer interaction [Murata 1998, Lank 2007, Ziebart 2010] and, more recently for automotive user interfaces. Ahmad [2014] proposed a system that predicts the intended icons on the interactive display on the dashboard early in the pointing gesture. He reported improved performance of the pointing task by reducing the target selection time using a particle filter-based target prediction system with which the user was able to select icons on screen before his hand physically touched the screen [Ahmad 2016]. They also reported a reduction in workload, effort, and duration of completing on-screen selection tasks. Biswas [2013; 2014] reported neural network-based prediction algorithms to locate intended target, while Lank and Phillip [Lank 2007, Pasqual 2014] proposed a method to predict endpoint using motion kinematics. Biswas [2017b] also presented an intelligent finger tracking system to operate secondary tasks in cars.

## 4. Wearable devices

Wearable devices are investigated in multiple disciplines of computer science like ubiquitous computing, Internet of Things (IoT), Assistive and Ambient Technology, and so on. Steinberger identified requirements for wearable devices in automotive as robust to body movement and capable of real-time data streaming [Steinberger 2017]. The data collection systems should be robust to movement artefacts associated with driving as it increases noise in data [Stern 2001, Baguley 2016]. Though reliable physiological measurement devices are heavyweight to wear and not suitable for automotive consumers [Liang 2007], advanced physiological measures are embedded in wearable devices like fitness trackers and smartwatches. Fitness trackers are capable of measuring heart rate activity and provide information regarding the user's seating status. Smartwatches measure biometric data, which is an indicator for driver drowsiness [Aguilar 2015]. Patterns are recognised from these devices' data to determine and characterise driver skills [Zhang 2010]. Wearable devices are used for acquisition of driver arousal data as an indicator of task engagement [Yerkes 1908]. Harman Becker automotive systems patented a head position monitoring device using a wearable loudspeaker that is worn on upper part of the body and a distance away from ears [Woelfl 2020]. Though consumers hesitate to use wearable technologies while driving, they still use wearable technologies for personal assistance like fitness wristbands [Kundinger 2020] which establishes a new market for acceptance of wearable technologies in everyday lives.

## 5. Speech Recognition and Non-visual interfaces

Speech recognition was investigated for reducing visual and manual distraction of drivers [McCallum, 2004; Sodnik, 2008]. Although automatic speech recognition works accurately in indoor environment, accuracy reduces in noisy environments inside a car [Gong, 1995; Baron, 2006]. The single-mode speech-based interaction has both positive and negative effects which are still debatable due to its controversial results from studies [He, 2014; Lee, 2001; Hua, 2010; He, 2015; McCartt, 2006]. A recent study showed that speech-based interaction in cars could cause distraction as much as drunk driving [He, 2015].

Researchers also attempted to eliminate or reduce visual search using gesture recognition techniques, but the systems either require to remember a set of gestures (AirGesture System

[May 2014]) or relative positions of screen items (BullsEye system [Weinberg 2012]). Additionally, such systems worked inferior to a touchscreen system in terms of driving performance or secondary task. It may be noted that, systems and services developed for elderly or disabled people often finds useful applications for their able bodied counterparts – a few examples are mobile amplification control, which was originally developed for people with hearing problem but helpful in noisy environment, audio cassette version of books originally developed for blind people, standard of subtitling in television for deaf users and so on. Considering these facts, technologies developed for users with visual impairment can have potential for non-visual interaction with IVIS system. A plethora of research has been conducted on navigation applications for blind users [Ganz 2011, Mulloni 2011, Rocha 2020]. Gorlewich [2020] and Palani [2020] formulated a set of guidelines for haptic and tactile rendering of touchscreen elements while ISO/TC 159/SC 4/WG 9 is working on a general purpose ISO standard on haptic and tactile interaction. It will be challenging to design an IVIS with 4mm inter-element spacing based on Gorlewich's [2020] study, however, it may also be noted that Prabhakar [2020] designed an automotive Head Up Display with a minimum inter-element spacing to make it accessible through eye gaze and gesture-based interaction and the interactive HUD improved driving performance with respect to traditional touchscreen in driving simulation study.

## 6. Gaze controlled interface

Research on eye gaze tracking started even in late 18th century although controlling an electronic inter-face using eye gaze or finger movement is relatively new concept. For eye gaze controlled interface, Zhai [1991] pioneered in MAGIC system, which aimed to use eye gaze track directly to improve pointing, in particular homing on a target using mouse. Ashdown and colleagues [2005] addressed the issue of head movement while tracking eye-gaze in a multiple monitor scenario. They used head tracking to switch pointer across screens, which was preferred by participants, but in effect increased pointing time. Dostal and colleagues [2013] addressed similar issues by detecting which monitor the user is looking at through analysing webcam video. The Sideways system [Zhang 2013] even eliminates personalized calibration and can scroll contents of a display screen by detect-ing eye gaze. The system identifies whether users are looking at the middle or sides of a display and if they are looking to the sides, the system scrolls content at the middle. Both Dostal's system and Sideways sys-tem does not use precise x and y coordinates to a move a mouse pointer. Fu and Huang [2013] pro-posed an input system hMouse, which moves a pointer based on head movement. They detected head movement by analysing video input and their system is found to outperform another similar system called CameraMouse [2013]. Fejtova's [2009] Magic Key system also uses a webcam like CameraMouse but the pointer is moved in screen based on position of nose (nosetrills to be precise). Selection is done by eye blinks. Bates [1999] multimodal eye tracking system allows zooming portion of screen using a polhemus tracker. Zandera and colleagues [2010] combine a BCI system with eye-gaze tracking, where EEG generated by imagining a rinsing action is trained to make a selection. However, their system had limited success in reducing pointing times. Penkar [2012] and colleagues investigated different dwell time durations for selecting target for an eye gaze controlled interface, although their study involved only selecting target at the middle of the screen. Later they extended the study [Lutteroth 2015] for selecting hyperlinks in a webpage but dwell time based selection would be inappropriate in automotive environment as it requires operators staring away from road for selecting target. Pfeuffer and Voelker explored fusing touchscreen interaction with gaze controlled system by using eye gaze for object selection and touch interaction for object manipulation. Pfeuffer [2016] explored desktop computing tasks like image searching and map navigation while Voelker [2015] investigated multi-screen display, which is more advanced in terms of coordinate mapping between horizontal and vertical displays compared to Dostal's [2013] system. However, our proposed work uses eye gaze for not only object selection but also for cursor movement.

Gaze controlled interfaces were already patented for automotive environment. For example, Mondragon and colleagues [2013] presented an eye gaze controlled smart display for passengers of vehicles. Users may point and select icons on the display by staring at appropriate portion of the screen. However, our proposed system is intended to be used by drivers and selection of target has to be faster than dwelling or staring away from road. Seder and colleagues [2012] presented a graphic projection display for drivers showing objects in road ahead. Users can select object on the projection display using different input modalities including eye gaze. However, the patent does not address a method of improving accuracy of the gaze tracking itself and it does not intend to operate the dashboard as the proposed system. Poitschke's [2011] study compared gaze controlled dashboard with touchscreen and reported higher reaction time for gaze controlled dashboard. Existing patents on eye gaze tracking mainly concerned about developing eye gaze tracking hardware [Voronka 2001] and using it to plot users' gaze location in a two-dimensional screen [Milekic 2009]. There are a set of patents which proposes to modify part of display based on eye gaze of users [Jacob 2013; Farrell 2005] or modifying rendering of a web browser [Martins 2003]. Farrell [2005] proposed to expand targets and proportionally contract other part of display based on eye gaze track of users although their system used a separate pointing device for cursor control instead of eye gaze. Vahtola [2014] and Jang [2012] proposed eye gaze enabled touch screen system where eye gaze tracking is used to confirm the touch input. The eye gaze tracking system required another cursor tracking program to operate. Pfeuffer and Voelker explored fusing touchscreen interaction with gaze controlled system by using eye gaze for object selection and touch interaction for object manipulation. Pfeuffer [2016] explored desktop computing tasks like image searching and map navigation while Voelker [2015] investigated multi-screen display.

## 7. Summary

The increasing number of functionalities influence the complexity of operating HMIs. Early automotive HMIs evolved in a non-user centric design approach. Present automotive HMIs improved the user experience by integrating new functionalities but added complexity to the interaction. This complexity causes inattentional blindness and leads to driver distraction. Existing research mostly explored infrared-based sensors in automotive environment for implementing virtual touch systems and least considered other options offering virtual touch. Wearable technologies are mostly explored for physiological measurement but not as a direct controller of user interfaces inside car. A plethora of target prediction technologies have already been tested with driving simulators but their accuracy often hampered by the sensing technology used inside an actual car. The tactile and haptic feedback systems are promising but needs more investigation for integration to automotive environment and display versatile content of an existing IVIS. Eye gaze controlled interfaces provide a viable alternative but most existing commercial eye gaze trackers ae not automotive compliance. Operating head down displays with eye gaze tracker has potential to increase duration of eyes off road while operating HUD may introduce he accommodation-vergance problem related to most augmented reality (AR) systems. Farrell [2005] noted that "humans use their eyes naturally as perceptive, not manipulative, body parts. Eye movement is often outside conscious thought, and it can be stressful to carefully guide eye movement as required to accurately use these target selection systems". However, automotive user interfaces do not require continuous manipulation of an on-screen pointer like graphical user interfaces in desktop computing. Unless a particular interaction is very familiar to the driver (like reaching for the gearbox while driving), he has to glance at the user interface. An accurate gaze tracking with target prediction technology can leverage this glance for pointing. Existing HMIs in commercial vehicles still rely on physical touch based interface with speech recognition technologies making good progress while high end vehicles investigated and deployed gesture recognition and eye tracking technologies. It will be important for researchers to externally validate any new technology inside a real car taking care of change in vibration and illumination. Novel technologies may also be tested with first seat passengers of semi-autonomous vehicle where the car can share the driving task with human occupants.